# Suppression of electron spin decoherence of the diamond NV center by a transverse magnetic field

Chang S. Shin[1,2,‡], Claudia E. Avalos[1,2], Mark C. Butler[1,2,‡‡], Hai-Jing Wang[1,2], Scott J. Seltzer[1,2], Ren-Bao Liu[3], Alexander Pines[1,2], and Vikram S. Bajaj[1,2,*]

[1]*Materials Sciences Division, Lawrence Berkeley National Laboratory, Berkeley, California 94720, USA*
[2]*Department of Chemistry and California Institute for Quantitative Biosciences, University of California, Berkeley, California 94720, USA*
[3]*Department of Physics and Centre for Quantum Coherence, The Chinese University of Hong Kong, Shatin, New Territories, Hong Kong, China*

## Abstract

We demonstrate that the spin decoherence of nitrogen vacancy (NV) centers in diamond can be suppressed by a transverse magnetic field if the electron spin bath is the primary decoherence source. The NV spin coherence, created in "a decoherence-free subspace" is protected by the transverse component of the zero-field splitting, increasing the spin-coherence time about twofold. The decoherence due to the electron spin bath is also suppressed at magnetic fields stronger than ~25 gauss when applied parallel to the NV symmetry axis. Our method can be used to extend the spin-coherence time of similar spin systems for applications in quantum computing, field sensing, and other metrologies.

In coupled spin systems, coherent superposition of spin states evolving in the presence of magnetic noise lose their spin coherence due to locally fluctuating magnetic fields [1–3]. Long coherence times are desirable because they permit the system to be interrogated for longer time, yielding improved sensitivity or precision in the estimation of frequencies. Achieving a long spin-coherence time is therefore a central problem in a variety of applications, such as magnetic resonance spectroscopy, magnetometry, and spintronics [4–9].

Recently, the negatively-charged NV center in diamond has attracted significant attention due to its unique optical and spin properties [10–15]. The NV center consists of a substitutional nitrogen atom at a carbon lattice site and a vacancy that is adjacent to the nitrogen atom. The ground state of the negatively charged NV center has electron spin $S = 1$, with $m_s = 0$ and $m_s = \pm 1$ sub-levels that are separated by a zero-field splitting of 2.87 GHz, as shown in Fig.1(a). The quantization axis of the NV center is along the $C_{3v}$ symmetry axis of the defect [16,17]. The electron spin of the NV center can be highly polarized to the $m_s = 0$ state via optical pumping [18], and its spin state-dependent fluorescence intensity allows convenient optical read-out of the spin state of the electron spin with high fidelity at room temperature [19].

Several approaches have been developed to extend the NV-electron spin coherence time, including the engineering of a spin system devoid of magnetic couplings through the use of magnetically pure ($^{12}$C) diamond; dynamical decoupling of the probed spin from the noisy environment through multiple-pulse sequences, such as Carr-Purcell-Meiboom-Gill (CPMG) or Uhrig dynamical decoupling (UDD) [20,21]; or the active detuning of bath spin fluctuations with a large external magnetic field [22]. All of these methods are effective in protecting spin coherence of the NV center from noisy environments and in doing so, they enhance the coherence times. For example, an electron-spin coherence time of several milliseconds at room temperature [12] and about half a second at 77 K [23] was achieved for NV centers in isotopically-pure $^{12}$C diamond. As a result, the NV center has been proposed as an ideal platform for the development of quantum-information processors and ultra-sensitive magnetometers.

It has been previously reported that, when the NV center is coupled to a mesoscopic quantum bath consisting of randomly distributed $^{13}$C nuclei in diamond [12,24,25], a spin coherence between $m_s = +1$ and $m_s = -1$ states (rather than between $m_s = 0$ and $m_s = \pm 1$) has an anomalously long lifetime [24,26]. This differential relaxation can be exploited by multiple pulse dynamical decoupling sequences. Here we report a complementary approach that corresponds, instead, to the creation of a long-lived state or decoherence-free subspace: the energy eigenstates of the NV spins are prepared such that the effective magnetic noise which causes fluctuation of the spin states is 'renormalized' by the transverse external magnetic field in combination with the transverse component of the zero-field splitting parameter $E$ of the NV center. The coherent superposition of the energy eigenstates thus becomes robust with respect to the magnetic noise. We theoretically and experimentally show that this method can be realized for an ensemble of NV centers in diamond. Specifically, experimental results showed that the coherence time was enhanced by a factor of ~2 at ~25 gauss. In principle, based on our theoretical model, the coherence time of the electron spin of the NV center can be enhanced by a factor of ~60, however other effects, such as decoherence due to the strongly coupled neighboring $^{13}$C nuclei or due to the electric field noise, limit the attainable enhancement [27,28].

In extremely pure diamond with a low substitutional paramagnetic nitrogen concentration of the order of a part per billion (ppb) (type II-a diamond), the coherent dynamics of the NV center depend strongly on the anisotropic hyperfine interaction with nearby $^{13}$C nuclei, randomly distributed in the diamond lattice at natural abundance (~1.1%) [10]. In spin-echo spectroscopy, the coherence of the NV center decays as the strongly-coupled $^{13}$C nuclear spins in the bath evolve in the presence of an effective field that depends on the electron spin-state of the NV center. With the external magnetic field applied along the NV quantization axis, all $^{13}$C nuclei have the same Larmor frequency when the electron is in $m_s = 0$ state. The phase accumulation associated with couplings between the electron and $^{13}$C nuclear spins is refocused by the spin echo if the two equal periods of the free evolution are a multiple of the Larmor period [10,27]. When the external magnetic field is misaligned from the NV quantization axis, however, transverse components of the magnetic field mix the electron spin states, and the effective $g$-tensor of a $^{13}$C nuclear spin (the Knight shift [10,13,27,29]) is enhanced in a way that depends on the position of each nucleus, relative to the NV center. The ensemble $^{13}$C nuclei in the bath are thus not uniformly coupled to electrons in the $m_s = 0$ state, which leads to fast decay of the electron-spin coherence and strong angular dependence of the coherence time of the NV center, with a minimum at $\theta = 90°$ (where the polar angle $\theta$ is measured between the NV axis and the external static magnetic field). This angular dependence has been extensively studied [10,27,30,31].

In diamond with a high concentration (~100 ppm) of paramagnetic nitrogen (P1 center) that can be categorized as type I-b diamond, the electrons form a strong spin bath, as schematically shown in Fig. 1(b). The spin-coherence of the NV center decays in the bath's fluctuating dipole field, which has a mean strength of ~1 MHz [22,32]. Each P1 center has strong hyperfine coupling (~100 MHz) between the electron spin and the $^{14}$N nuclear spin. At zero field, the energy eigenstates are doubly-degenerate entangled states between the electron and nuclear spins of the P1 center, and the dipolar interaction between P1 centers induces rapid random reorientation of the P1 spins. As the external magnetic field is increased, this spin-reorientation becomes improbable, due to the large energy difference between eigenstates as compared to the strength of the dipolar interaction. This slows down fluctuations of the bath spins, resulting in suppression of decoherence of the NV center. For example, in a strong external magnetic field of 740 gauss applied along the NV axis, the coherence time of a single NV center was an order of magnitude longer than at zero magnetic field [22]. Here we instead study the coherence time of ensemble NV centers in type I-b diamond in a weak external magnetic field ($B_0 \leq$ 75 gauss). We report sharp angular dependence of the coherence time with a maximum at $\theta = 90°$, in contrast to the case where $^{13}$C nuclei dominate the bath dynamics.

In what follows, hyperfine interaction between electron spins and $^{14}$N nuclear spins of the NV center can be safely neglected, since spin-spin relaxation, induced by the hyperfine interaction, is largely suppressed due to the energy mismatch. The Hamiltonian of the NV center can be written as [16,33]

$$H = D\left[S_z^2 - \frac{S(S+1)}{3}\right] + \gamma[\mathbf{B} + \mathbf{b}(t)] \cdot \mathbf{S} + E\left(S_x^2 - S_y^2\right), \qquad (1)$$

Where $D$ and $E$ are the parallel and transverse zero-field splitting parameters of the NV center, respectively. Note that the coupling constant $D$ characterizes spin-spin interactions, while $E$ is due to the crystal field induced by the non-axial components of the strain that lower the symmetry [34]. $\gamma/2\pi \approx$ 2.8 MHz/gauss is the gyromagnetic ratio, $\mathbf{S}$ is the vector operator for the electron spin, $\mathbf{B}$ is the external static magnetic field, and $\mathbf{b}(t)$ is the fluctuating dipole field experienced by the NV center due to the bath.

In a weak magnetic field, i.e., $D \gg |\gamma \mathbf{B}|, |\gamma \mathbf{b}(t)|$, terms in the Hamiltonian that couple states separated by the large zero-field splitting can be treated as perturbations. To do so, we write an effective Hamiltonian that includes elements only within the space spanned by the nearly-degenerate states, i.e. $m_s = \pm 1$ states. The effective Hamiltonian can be approximated as a series [35,36], with eigenstates defined as $|0\rangle, |+\rangle$ and $|-\rangle$; the latter two states arise from mixing between $m_s = \pm 1$ states caused by the transverse component of the magnetic field. From second-order perturbation theory, the transition energies are approximated as

$$E_{|0\rangle \leftrightarrow |\pm\rangle} = D + \frac{3\delta E}{2} \pm \left[ \gamma^2 (B_z + b_z)^2 + |\widetilde{E}|^2 \right]^{1/2}, \qquad (2)$$

where $\delta E$ and $\widetilde{E}$ are defined as

$$\delta E = \frac{\gamma^2 \left[ (B_x + b_x)^2 + (B_y + b_y)^2 \right]}{D}, \qquad (3)$$

$$\widetilde{E} = E + \frac{\gamma^2 \left[ (B_x + b_x) \pm i(B_y + b_y) \right]^2}{2D}. \qquad (4)$$

Comparison of exact energies with energies calculated using second-order perturbation theory shows that for typical values of the parameters appearing in Eqs. (2) through (4), second-order perturbation theory is valid for describing frequency-shifts in a perpendicular magnetic field of up to ~150 gauss.

When $E/2\pi$ is on the order of several MHz and $|\gamma \mathbf{b}(t)|/2\pi \sim 1$ MHz, the radical in Eq. (2) can be expanded as a series in $b_z$, and we find that the transition frequencies are

$$E_{|0\rangle \leftrightarrow |\pm\rangle} \approx D \pm \left( \gamma^2 B_z^2 + |\widetilde{E}|^2 \right)^{1/2} + \gamma \widetilde{b}(t), \qquad (5)$$

where the renormalized noise $\gamma \tilde{b}(t)$ is defined as

$$\gamma \tilde{b}(t) \equiv +\frac{3\delta E}{2} \pm \frac{\gamma^2 B_z b_z}{\left(\gamma^2 B_z^2 + |\tilde{E}|^2\right)^{1/2}} \pm \frac{|\tilde{E}|^2 \gamma^2 b_z^2}{2\left(\gamma^2 B_z^2 + |\tilde{E}|^2\right)^{1/2}} + O\left(b_z^3\right). \qquad (6)$$

In the case where $B_z$ is nonzero but $B_x = B_y = 0$, the renormalized noise can be approximated as

$$\gamma \tilde{b}(t) \approx \pm \frac{\gamma^2 B_z b_z}{\left(\gamma^2 B_z^2 + |\tilde{E}|^2\right)^{1/2}} \pm \frac{|\tilde{E}|^2 \gamma^2 b_z^2}{2\left(\gamma^2 B_z^2 + |\tilde{E}|^2\right)^{1/2}} + O\left(b_z^3\right), \qquad (7)$$

and the largest contribution of the noise to the frequency shift between the energy eigenstates is from the first term in Eq. (7), i.e.

$$\gamma \tilde{b}(t) \approx \frac{\gamma^2 B_z b_z}{\left(\gamma^2 B_z^2 + |\tilde{E}|^2\right)^{1/2}}. \qquad (8)$$

On the other hand, in the case where $B_x$ is nonzero but $B_y = B_z = 0$, the linear noise term $B_z b_z$ in Eq. (6) becomes zero, and the renormalized noise can be approximated as

$$\gamma \tilde{b}(t) \approx \frac{2\gamma^2 B_x b_x}{D} \pm \frac{\gamma^2 b_z^2}{2|\tilde{E}|}. \qquad (9)$$

The frequency-shift given by Eq. (9) includes a term quadratic in $b_z$ as well as a term linear in $b_x$ that is scaled by the zero-field splitting $D$. Equations (8) and (9) show that when $|\gamma \mathbf{b}(t)| \ll |\gamma \mathbf{B}| \ll D$, the transverse magnetic field reduces the effective noise amplitude, $\tilde{b}(t)$, experienced by the electron spin of the NV center, suppressing the mechanism of decoherence. For example, numerical calculation using parameters $D/2\pi = 2870$ MHz, $E/2\pi = 4.85$ MHz, and $\gamma b_x/2\pi = \gamma b_y/2\pi = \gamma b_z/2\pi \sim 0.7$ MHz (estimated from the free induction decay of our sample with estimated NV concentration of ~ 5 ppm and a nitrogen concentration of < ~100 ppm) shows that the frequency shifts associated with magnetic noise are smaller by about a factor of ~8 when a magnetic field of 25 gauss is applied perpendicular to the NV quantization axis, as compared to the case where the magnetic field is applied parallel to the NV axis.

We experimentally measured the coherence decay of the NV center using a Hahn Echo pulse sequence with optically detected electron spin resonance (ODESR) techniques at a static magnetic field of ~25 gauss, applied parallel or perpendicular to the NV axis. The detailed experimental procedure for ODESR is described elsewhere [37]. As shown in Fig. 2, the spin-echo signal of the NV center decays about two-fold more slowly when the field is perpendicular to the NV axis. We note that Dolde et al. showed that the coherence time of a single NV center in the presence of inhomogeneous broadening is greatly increased when the longitudinal magnetic field becomes zero [28], which could be explained by the renormalization of the linear noise term shown in Eqs. (8) and (9).

Next, the coherence time of the NV center at various polar angles, $\theta$, was measured, and a striking angular dependence of the coherence time on the external magnetic field was observed. Specifically, the electron-spin coherence time of the NV center is relatively insensitive to the angle ($\theta$) between the NV quantization axis and the external magnetic field when $\theta$ is varied between 0° and ~80°, but $T_2$ increases sharply as the angle is increased from 85° to 90°, with a maximum at $\theta = 90°$, as shown in Fig. 3.

This angular dependence of the spin coherence time is due to renormalization of the noise, and the mechanism is completely different than in the case where $^{13}$C nuclear spins dominate the bath. In a weak magnetic field, the spin-coherence time of the NV center in a bath dominated by the $^{13}$C nuclear spins is very sensitive to the magnetic field alignment and is at a minimum with the external magnetic field perpendicular to the NV axis. This is due to position-dependent modification of the effective g-tensor of nearby $^{13}$C nuclear spins when the transverse component of the external field enhances the mixing of electronic and nuclear states [10,27,30]. In contrast, the coherence time of the electron spin is lengthened by a transverse field when the NV center interacts with a bath of P1 centers, since the first-order frequency shift associated with the fluctuating field component, $b_z$, is eliminated.

In general, fluctuating magnetic field components along *x*, *y* and *z* can cause decoherence of the NV electron spins by inducing inhomogeneous frequency shifts. However, direct spin-flip relaxation due to the perpendicular components of the fluctuating magnetic fields is negligible, since the spectral density of the fluctuating dipole fields that couple the electron spin states (separated by ~7 MHz at the transverse field of ~25 gauss) is at least three orders of magnitude smaller than the secular broadening. The coherence time is limited by secular broadening, and it can be written as [2,33]

$$\frac{1}{T_2} = \gamma^2 \tilde{b}^2 \tau_0, \tag{10}$$

where $\tau_0$ is the characteristic correlation time of the fluctuating field in the bath. On the basis of Eq. (10), we expect that the coherence time ($T_2$) will increase greatly due to the aforementioned noise scaling. For instance, a decrease in magnetic noise by a factor of ~8 corresponds to an increase in the coherence time by a factor of ~60 at 25 gauss field applied perpendicular to the NV axis.

Though the theoretical model predicts a large enhancement in the coherence time, our experimental results might be limited by residual linear noise of the form given by Eq. (8), due to the imperfect zeroing of $B_z$. For example, with residual $B_z \sim 0.5$ gauss, comparable to the Earth's magnetic field (the ambient field in our experiment), the linear noise term with $B_z b_z$ in the numerator dominates the renormalized noise in Eq. (6). It is also possible that the experimental results might be limited by other sources of decoherence not included in the model, such as electric field noise, strain noise and/or the strongly coupled $^{13}$C nuclei [27,28,31]. A larger enhancement of spin-coherence time is therefore expected for NV centers in isotopically enriched $^{12}$C diamond, where decoherence, in the absence of $^{13}$C nuclei, is primarily due to the residual P1 centers [12,32].

Under the assumption that the term proportional to $B_z b_z$ in Eq. (6) is dominant, experimental measurements of $(1/T_2)$ as a function of polar angle were fitted to the Eq. (10) using $\tau_0$ as a free parameter with $\gamma b_x/2\pi = \gamma b_z/2\pi \sim 0.7$ MHz. We also include an offset term that accounts for the residual noise contributions from other terms in Eq. (6). Excellent agreement between experimental data and the theoretical model was obtained, as shown in Fig. 3. At $\theta = 55$ or $65°$, transition frequencies of the NV centers in other orientations are close to the transition frequencies of the centers we intended to observe. Therefore, outliers in red at $55°$ and $65°$ are not included in the fitting, since the coherence decay was likely affected by efficient cross-relaxation between the NV centers in different orientations or by off-resonant contributions of those NV centers. This would result in faster decay due to imperfections in the $\pi$ pulse for the off-resonant components.

We further measured the coherence time of the NV center as a function of the magnitude of the external magnetic field, oriented either parallel to the NV axis or perpendicular to the NV axis, as shown in Fig. 4(a). As the external magnetic field parallel or perpendicular to the NV axis increased to ~ 20 gauss, the coherence time of the NV center increased rapidly. At zero field, there are degenerate states in a P1 center, allowing for fast transitions driven by the dipole fields. However, the degeneracy is removed as the field is increased to ~25 gauss, where the electron Zeeman interaction is comparable to the hyperfine interaction of the P1 center. This freezes out most of the dynamics of the bath, which is gradually suppressed more fully as the field increases further [22]. When a larger magnetic field (> ~25 gauss) was applied parallel to the NV axis, the NV coherence time slowly increased. On the other hand, the NV coherence time increased faster when the magnetic field was applied perpendicular to the NV axis, which could be due to the renormalization of the noise through the $\widetilde{E}$ term in Eqs. (4) and (9). We note that in principle further suppression of decoherence can be achieved by applying transverse electric and/or strain fields that can increase $\widetilde{E}$ through the electric dipole interaction [28]. We found that the dependence of coherence time ($T_2$) on the perpendicular magnetic field levels off at larger magnetic fields of ~50 gauss. This is possibly due to the non-negligible contribution of the linear noise term in Eq. (9).

Just as in the spin echo experiments, the coherence times under the CPMG pulse were also enhanced by a transverse field, i.e. to 9.8 μs from ~5.8 μs with a longitudinal field, as shown in Fig. 4(b). Spin coherence time under dynamical decoupling pulse has been reported to observe a power-law relation $T_2 \sim N^r$, where $N$ is the number of π-pulse, and the index $r$ is related to characteristics of the bath [32,38,39]. The spin coherence time of the ensemble of NV centers in our sample fits well to the power-law relation with *r* ~ 0.47±0.04 for longitudinal field and *r* ~ 0.42±0.01 for transverse.

In conclusion, we demonstrated that transverse magnetic field enhanced the spin coherence times by ~ a factor of 2 from that with longitudinal magnetic field. Experiments qualitatively agreed with our noise renormalization model that shows effective dipole fields experienced by the NV electron spins can be renormalized by the external magnetic field and the zero-field splitting parameters of the NV center, resulting in suppression of the decoherence of the NV electron spin at a given external magnetic field. Our measurements also showed that the spin coherence time of the NV center in type I-b diamond is associated with fast spin reorientation of the P1 center, which are greatly suppressed in magnetic fields larger than ~25 gauss. Further enhancement of spin-coherence time with transverse magnetic field is expected for NV centers in isotopically enriched $^{12}$C diamond, where $^{13}$C nuclei are greatly depleted. Our method can be used to enhance the coherence time of NV centers and similar spin systems for various applications in metrology and quantum information technology.

**Acknowledgments**

This work was supported by the Director, Office of Science, Office of Basic Energy Sciences, Materials Sciences and Engineering Division, of the U.S. Department of Energy under Contract No. DE-AC02-05CH11231. The authors thank Ran Fischer and Dmitry Budker for useful discussions.

‡Current address: Halliburton Energy Services, Inc., 3000 N Sam Houston Pkwy. E, Houston, TX 77032, USA

#Current address: William R. Wiley Environmental Molecular Sciences Laboratory, Pacific Northwest National Laboratory, Richland, WA 99352, USA

*vikbajaj@gmail.com

**Figure captions**

**FIG. 1.** (Color online) **(a)** Energy level structure of the NV center showing spin triplet ground ($^3A_2$) and excited ($^3E$) states with single states (green arrows for optical excitation at 532nm, red arrow for fluorescence decay, grey arrows for non-radiative decay via singlet states, and blue arrow for microwave excitation). **(b)** Schematic of an NV center spin, coupled to bath spins dominated by substitutional nitrogen (P1) spins (blue balls with arrow for P1 spins).

**FIG. 2.** (Color online) Coherence decay of ensemble NV electron spins with an external magnetic field of ~25 gauss applied parallel or perpendicular to the NV quantization axis (solid lines are fits to a single exponential decay). The measured coherence time was 1.17±0.01 µs for $\theta = 0°$ (crosses) and 2.01±0.01 µs for $\theta = 90°$ (circles).

**FIG. 3.** (Color online) Coherence decay rate ($1/T_2$) as a function of magnetic field orientation ($\theta$) with respect to the NV axis at ~25 gauss. The solid line is a fit of experimental data ($1/T_2$) to Eq. (10) using $\tau_0$ as a free parameter, together with an offset that accounts for the residual noise contributions. The value $\gamma b_z/2\pi \sim 0.69 \pm 0.01$ MHz was estimated from the free induction decay, while $E/2\pi \sim 4.85$ MHz was estimated from the ODMR spectrum at ambient field. The outliers at 55° and 65°, which might be affected by NV centers in other orientations, were not included in the fitting.

**FIG. 4.** (Color online) **(a)** Spin-coherence time as a function of the amplitude of a magnetic field applied parallel to the NV axis ($\theta = 0°$), or perpendicular to the NV axis ($\theta = 90°$). (Solid lines

serve to guide the eye) **(b)** Spin-coherence time of ensemble NV centers as a function of the number of $\pi$-pulses ($N$) in the CPMG pulse sequence at ~75 gauss applied parallel or perpendicular to the NV quantization axis. Solid lines are fits to $T_2 \sim N^r$ with $r = 0.47 \pm 0.04$ for $\theta = 0°$ and $0.42 \pm 0.01$ for $\theta = 90°$.

# Figures

**FIG. 1.**

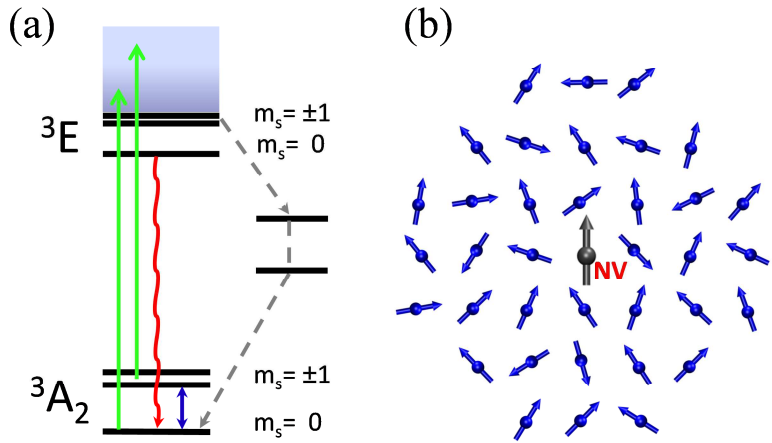

**FIG. 2.**

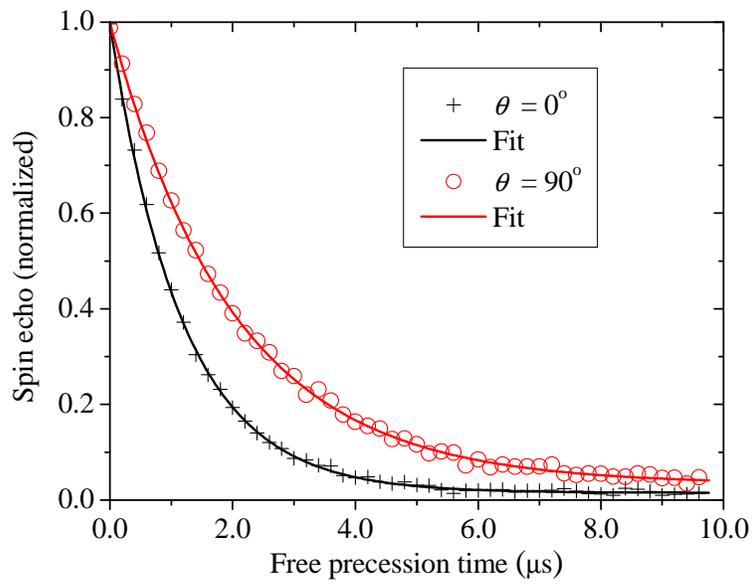

**FIG. 3.**

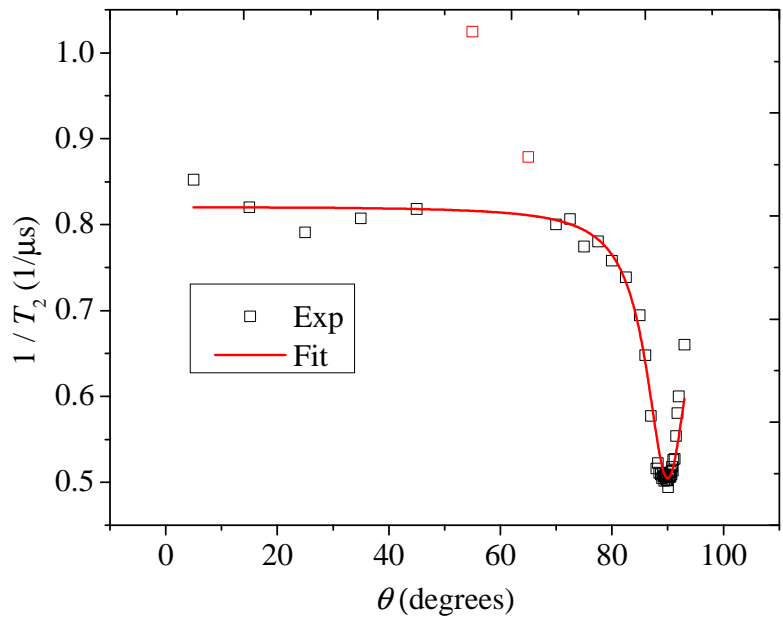

**FIG. 4**

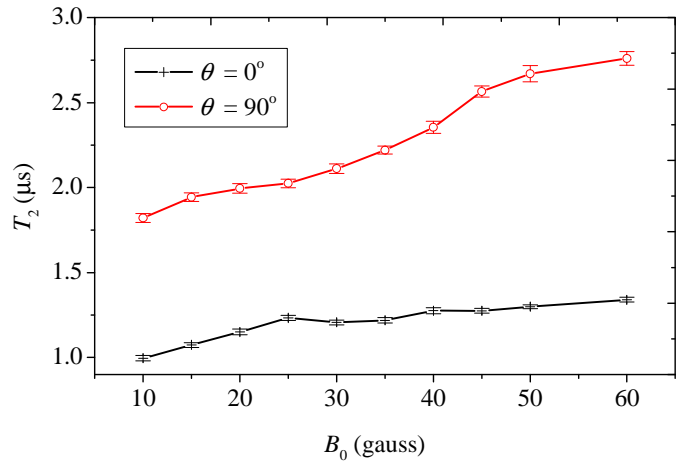

**(a)**

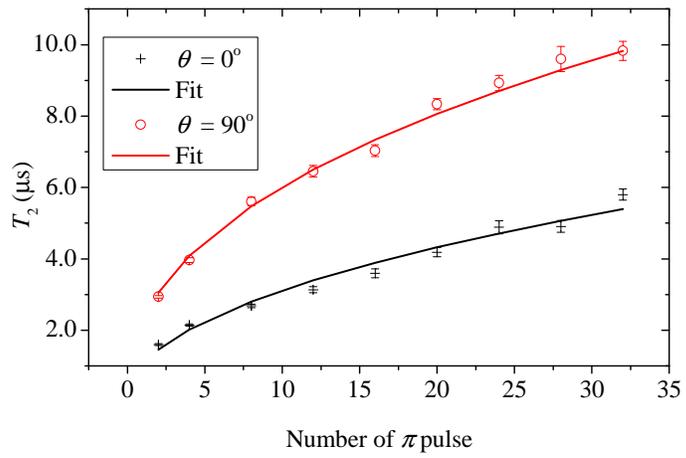

**(b)**

20